\documentclass[aps,prx,reprint,superscriptaddress,amsmath,amssymb,floatfix]{revtex4-2}

\usepackage{graphicx}
\usepackage{booktabs}
\usepackage{bm}
\usepackage[colorlinks=true, linkcolor=blue, citecolor=blue, urlcolor=blue]{hyperref}
\usepackage{braket}

\begin{document}

\title{All-Optical Neural Network by First-Photon Detection}

\author{Ishan Darji}
\author{Santosh Kumar}
\author{Yu-Ping Huang}
\email[Contact author: ]{yhuang5@stevens.edu}
\affiliation{Stevens Institute of Technology, Hoboken, New Jersey 07030, USA}
\affiliation{Center for Quantum Science and Engineering, Stevens Institute of Technology, Hoboken, New Jersey 07030, USA}

\date{November 25, 2026} 

\begin{abstract}
Optical implementations of computing, which can potentially exploit quantum degrees of freedom in addition to the massive parallelism of light, are being developed as an energy-efficient approach to machine learning. To this end, it is essential to understand and thereby utilize their fundamental advantages, particularly those stemming from the particle nature of light. We introduce a simple, single-layer nonlinear optical neural network (NLONN) based on sum-frequency generation (SFG), in which a programmable classical pump enables neural processing at the single-photon level. At the input layer, we have a coherent signal that carries the image written on it as a phase. In the trainable layer, a pump beam shaped into a superposition of Hermite--Gaussian (HG) modes drives SFG in a nonlinear crystal at the Fourier plane. The output layer collects the upconverted photons through a single-mode fiber and detects them using a single-pixel, visible single-photon detector. The pump-mode kernels are trained in situ from responses measured on training digits. Using the arrival time of the first detected photon, the NLONN classifies three MNIST digits with an accuracy of 56.2 $\pm$ 2.3\,\% on 450 test images. This outperforms a camera at the object plane (33.3\,\%, random guessing) and at the Fourier plane (41.3\,\%). It also outperforms a phase layer trained before the lens (51.5\,\%), while using only 54 trained parameters compared with 6,724. Its performance is comparable to that of a phase layer trained at the Fourier plane (57.7\,\%). We show that the NLONN's performance can be further improved through enhanced pump encoding and higher-order kernel-mode superpositions.

\end{abstract}

\maketitle

\section{Introduction}
\label{sec:intro}

The increasing computational demands of artificial intelligence (AI) are challenging the energy efficiency, scalability, and accessibility of conventional digital computing platforms. Optical neural networks (ONNs) are being researched to perform the linear operations that dominate machine learning~\cite{Wetzstein2020,Shastri2021,McMahon2023,Zhou2026ER}, using the massive parallelism they can provide at low energy cost. Such ONNs have been demonstrated using coherent nanophotonic meshes~\cite{Shen2017}, diffractive networks~\cite{Lin2018,Zhou2021}, integrated photonic tensor cores and convolution accelerators~\cite{Feldmann2021,Xu2021}, end-to-end on-chip classifiers~\cite{Ashtiani2022}, all-optical nonlinear activations~\cite{Zuo2019}, and architectures based on photoelectric multiplication~\cite{Hamerly2019,Sludds2022}. Training and running large models have a substantial energy cost~\cite{Strubell2019}, while optical energy required for linear operations can be small. Handwritten digits have been classified with less than one detected photon per multiplication~\cite{Wang2022}. Beyond parallelism, light also carries quantum degrees of freedom that can be exploited through measurement.

Advances in single-photon technologies are enabling new capabilities across a wide range of scientific and technological fields~\cite{Cheng2022,Wu2026,Li2022,Ma2025,Wang2023}. These capabilities are opening new opportunities for secure communication, quantum-enhanced sensing, low-light imaging, and photonic information processing. One fundamental limit of optical implementations asks for a decision after detecting a single photon~\cite{Kirmani2014}. A single detection event can bring information from several features only if the optics before it allow interference between the spatial modes shared by these features~\cite{Naruse2016}. The question for machine learning is then which trainable measurement allows us to extract this information about an image from a single photon. This connects optical computing with quantum measurements and quantum machine learning~\cite{Biamonte2017,Qiu2019,Ortolano2023}. The single-photon constraint has been analyzed theoretically for image classification~\cite{Fischbacher2021}. A ten-class classifier that learns only which pixel the photon reaches cannot exceed 22.96\,\% accuracy for MNIST digits. A unitary transform trained before detection, which allows a photon to self-interfere, can reach 41.27\,\% accuracy.

In this work, the unitary computation preceding detection is implemented through a nonlinear optical process. Specifically, in sum-frequency generation (SFG) using a periodically poled lithium niobate (PPLN) bulk crystal, a structured pump field drives the conversion of an input signal photon into an upconverted photon that is collected in a single output mode. The upconversion probability is proportional to the squared overlap of the photon with a mode defined by the pump. This mode-selective frequency-conversion technique is known as quantum pulse gating or quantum parametric mode sorting~\cite{Eckstein2011,Brecht2015,Shahverdi2017}. It has also been extended to spatial-mode selectivity~\cite{Zhang2019}, image processing~\cite{Kumar2019} and single-photon classification~\cite{Li2024,Li2024b}.

We build a nonlinear optical neural network (NLONN) on this principle and probe how far it reaches at the limit of a single detected photon. The image is phase-encoded on a coherent signal beam at the input layer using a spatial light modulator (SLM). A decision made from a detection without self-interference of this signal is a random guess. Trainable kernels are written on another coherent pump beam as a superposition of Hermite--Gaussian (HG) modes with another SLM. SFG in a bulk crystal acts at the Fourier plane of a 4f system after the beams are combined, and the upconverted light is collected in a single-mode fiber, which performs the mode projections, and read by a single-pixel detector. No spatial degree of freedom is left at the detector, so all spatial information must be extracted in the crystal. The kernels are trained in situ from measured responses by adjusting the coefficients of the pump's superposition modes. On a three-class task, we compare the NLONN at the first detected photon with linear-optical measurements built from fixed and trained phase elements. From the first detected photon, the NLONN reaches 56.2\,\% on MNIST digits 2, 4 and 7, against 41.3\,\% for a lens and camera, and with 54 trained parameters it exceeds a phase layer with 6,724. Its accuracy rises with the order of the pump modes. 

In the following, Sec.~\ref{sec:theory} provides the theoretical background of the NLONN. Section~\ref{sec:experiment} gives details of the experiment, and Sec.~\ref{sec:results} discusses the results. Section~\ref{sec:discussion} places the results in context alongside current limitations, and Sec.~\ref{sec:conclusion} concludes.

\section{Theoretical framework}
\label{sec:theory}

\begin{figure*}[t]
\centering
\includegraphics[width=\textwidth]{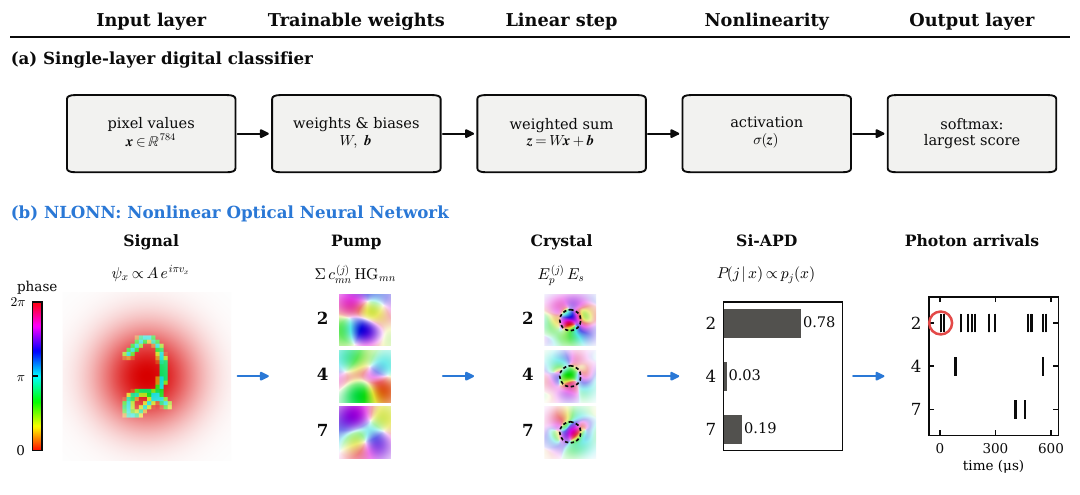}
\caption{Conceptual diagram illustrating a comparison between a digital neural network and the NLONN. (a) A single-layer digital classifier multiplies real pixel values by weights trained in software (for three classes, $3\times784$ weights and 3 biases), sums them, applies a deterministic activation and names the class with the largest score. (b) The NLONN, shown for one test image of digit `2' decided correctly in the experiment by its first recorded photon. \emph{Input layer:} the signal field on its modulator, $\psi_x\propto A\,e^{i\pi v_x}$; the digit is carried only in the phase, and the intensity is the same for every image. \emph{Trainable weights:} the pump superpositions of ten Hermite--Gaussian modes, trained in situ, one kernel per class (rows 2, 4, 7), drawn as the target fields written on the pump modulator; they have 54 free real parameters. \emph{Linear step:} mode projection of the signal onto the pump kernel within the crystal, Eq.~\eqref{eq:2}, computed at the center of the crystal from the ideal model. The black dotted circle marks the back-propagated mode of the single-mode fiber. \emph{Nonlinearity:} the detector's click applies the Born rule; the bars are each kernel's share of the first click, $p_j(x)/\sum_{j'}p_{j'}(x)$, from the measured photon counts of this image. \emph{Output layer:} the image's recorded clicks under each kernel, at their recorded times; the first click (red circle) names the class. In the field images, hue is the phase (color bar) and brightness corresponds to the field amplitude.}
\label{fig:1}
\end{figure*}

Figure~\ref{fig:1} illustrates a comparison between (a) a conventional digital neural network and (b) our NLONN in a layer-by-layer architecture. Both are single-layer networks: the NLONN has one trainable layer, the coherent projection set by the pump, between the encoded input and the detector output.

At the input layer, an image $x$ with gray levels $v_x(\bm r)\in[0,1]$ is phase-encoded on a coherent signal beam with transverse amplitude $A(\bm r)$, so that the signal photon carries the state

\begin{equation}
\psi_x(\bm r) = \mathcal N\,A(\bm r)\,e^{i\pi v_x(\bm r)} ,
\tag{1}
\label{eq:1}
\end{equation}

with $\mathcal N$ fixing the normalization, $\int|\psi_x|^2\,d^2r=1$. The intensity $\mathcal N^2 A(\bm r)^2$ is the same for every image. All the information is in the phase, and a detector that does not let the photon interfere with itself cannot reach it.

The trainable layer projects the photon by sum-frequency generation. The signal and a strong pump are focused into a $\chi^{(2)}$ crystal of length $L$ at the Fourier plane of their spatial light modulators. The upconverted sum-frequency light is collected in a single-mode fiber. The probability amplitude of a single photon producing a click is the projection of its state onto a mode set by the pump and the fiber,

\begin{equation}
\begin{split}a_j&\propto\int dz\,\big\langle \phi_f\,E_p^{(j)*}e^{-i\Delta kz}\,\big|\,\mathcal L_z\psi_x\big\rangle\\&=\Big\langle \underbrace{\int dz\;\mathcal L_z^\dagger\big[\phi_f\,E_p^{(j)*}e^{-i\Delta kz}\big]}_{K_j}\;\Big|\;\psi_x\Big\rangle .\end{split}
\tag{2}
\label{eq:2}
\end{equation}

Here $z$ runs over the crystal's length, and $\mathcal L_z$ carries the signal from its modulator through the first lens of the 4f system to depth $z$ in the crystal. The bracket is the overlap over the transverse plane at that depth, $E^{(j)}_p$ is the pump field in configuration $j$ there, $\phi_f$ is the fiber mode back-propagated into the crystal, and $\Delta k$ is the phase mismatch. Read from the inside out, $K_j$ is the kernel set by the pump and fiber modes. At each depth, it is the mode that the pump converts into the fiber mode. It is summed along the crystal and carried back to the signal modulator. Equation~\eqref{eq:2} is the overlap of the three fields in the crystal, $\int dz\int d^2\rho\,\phi_f^{*}E^{(j)}_pE_s\,e^{i\Delta kz}$ with $E_s=\mathcal L_z\psi_x$~\cite{Eckstein2011,Brecht2015,Kumar2019}. Propagation and the lens are linear in the field, so the signal photon's state separates from everything else. The pump is two orders of magnitude stronger than the signal, so we take it as undepleted. The conversion probability per photon is also far below one, so each signal photon upconverts separately and independently, and the probability that a photon in state $\psi_x$ produces a click is

\begin{equation}
p_j(x) = \eta\,\bigl|\langle K_j|\psi_x\rangle\bigr|^2 , \qquad K_j = \sum_{mn} c^{(j)*}_{mn}\,\kappa_{mn} .
\tag{3}
\label{eq:3}
\end{equation}

Here $\eta$ collects the conversion and detection efficiency, which accounts for photons that do not convert or do not produce a click. The pump is a superposition of HG modes, $E_p^{(j)}=\sum_{mn}c^{(j)}_{mn}\mathrm{HG}_{mn}$. It enters Eq.~\eqref{eq:2} linearly and conjugated. The $\chi^{(2)}$ process is nonlinear in the two optical fields and linear in the signal photon, so the photon's superposition is preserved through the conversion, and the weights enter through the pump field. The kernel is therefore the same superposition, with conjugated coefficients, of the modes $\kappa_{mn}$, where $\kappa_{mn}$ is the kernel of Eq.~\eqref{eq:2} for a pure $\mathrm{HG}_{mn}$ pump. The coefficients $c^{(j)}_{mn}$ are the trainable weights of the layer. Writing $g_{mn}(x)=\langle\kappa_{mn}|\psi_x\rangle$,

\begin{equation}
p_j(x) = \eta \sum_{mn}\sum_{m'n'} c^{(j)}_{mn}\,c^{(j)*}_{m'n'}\; g_{mn}(x)\,g^{*}_{m'n'}(x) .
\tag{4}
\label{eq:4}
\end{equation}

At the output layer, a silicon avalanche photodiode registers the upconverted photon. The click applies the Born rule, which is the network's nonlinear readout. A trained NLONN uses one kernel per class, kernel $j$ naming class $j$, each shown for the same time on the pump SLM. The first click names class $\hat c$ with probability
\begin{equation}
P(\hat c\,|\,x) = \frac{p_{\hat c}(x)}{\sum_{j} p_j(x)} ,
\tag{5}
\label{eq:5}
\end{equation}
and the accuracy is the mean of $P(y_x|x)$ over test images of true class $y_x$. Every image is classified on its first click. The input is a coherent signal beam, not a single-photon source. Because its photons are converted independently, the clicks under kernel $j$ form a Poisson process with rate $R_j(x)\propto p_j(x)$. The kernel whose click comes first names $\hat c$ with the probability of Eq.~\eqref{eq:5}, as the first converted photon of a single-photon input would. Equation~\eqref{eq:5} can therefore be estimated from recorded photon arrivals (Sec.~\ref{subsec:accuracy-from-recorded-photons}). Repeated single-shot trials on the same input state estimate the same probability with a smaller statistical error.

\begin{figure*}[t]
\centering
\includegraphics[width=0.8\textwidth]{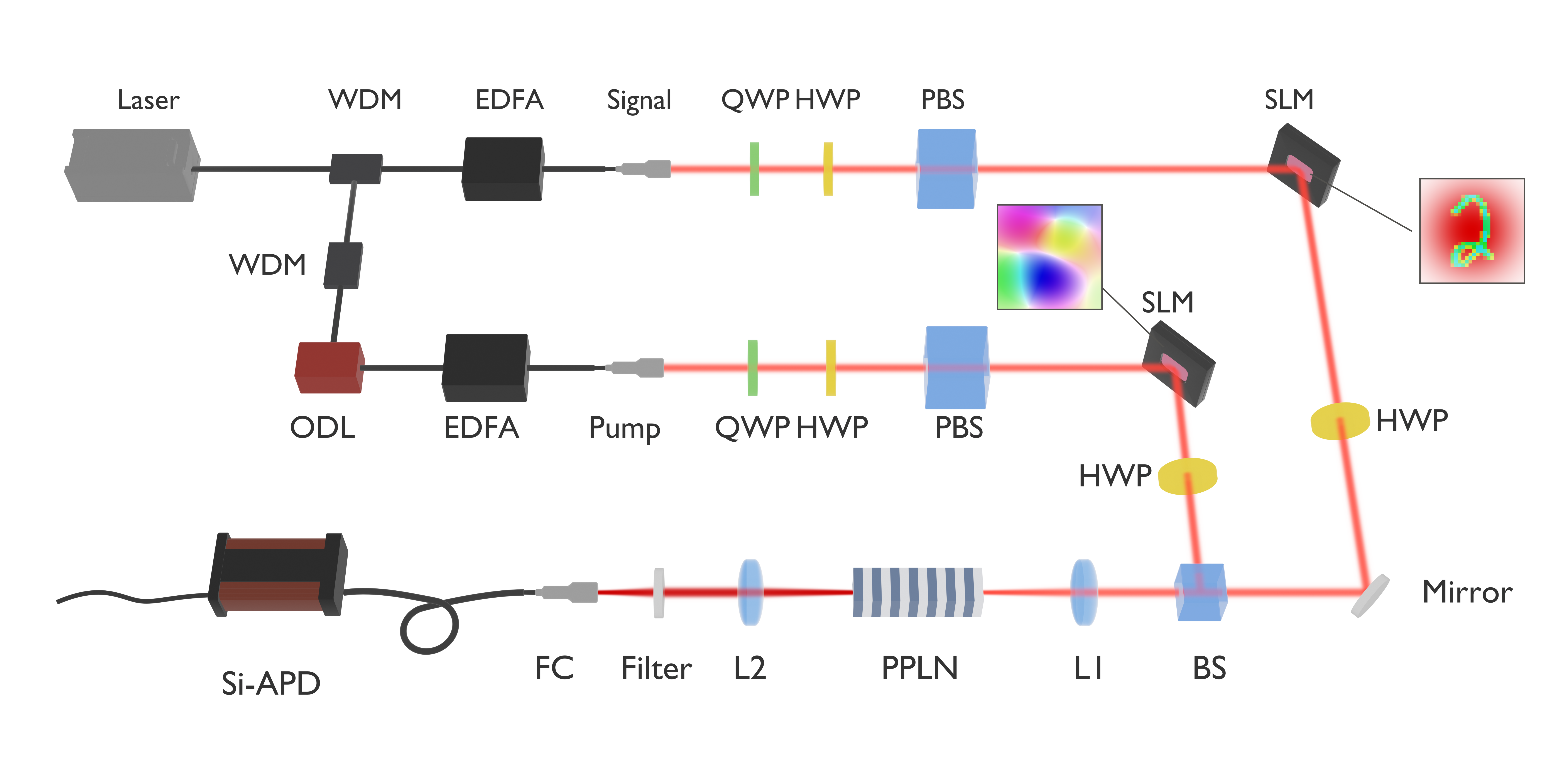}
\caption{The experimental setup. The signal SLM writes the digit as a phase (inset), and the pump SLM writes the kernel (inset). L1 and L2 form a 4f system with the SLMs at its object plane, the PPLN crystal at its Fourier plane and the fiber coupler at its image plane. Photon arrivals on the Si-APD are recorded by a Swabian Time Tagger Ultra (not shown). WDM, wavelength-division multiplexer; EDFA, erbium-doped fiber amplifier; ODL, optical delay line; QWP, HWP, quarter- and half-wave plates; PBS, polarizing beam splitter; SLM, spatial light modulator; BS, beam splitter; PPLN, periodically poled lithium niobate; FC, fiber coupler; Si-APD, silicon avalanche photodiode.}
\label{fig:2}
\end{figure*}

\section{Experiment}
\label{sec:experiment}

\subsection{Apparatus}
\label{subsec:apparatus}

Figure~\ref{fig:2} shows the experimental setup. A fiber mode-locked laser (100 MHz repetition rate) is demultiplexed into a signal channel at 1545.35 nm (1.17 nm wide) and a pump channel at 1558.57 nm (0.44 nm). Each arm has its own erbium-doped fiber amplifier, and the pump arm has an optical delay line that overlaps the two pulse trains in time at the crystal. On each arm a quarter-wave plate (QWP), a half-wave plate (HWP) and a polarizing beam splitter (PBS) set the optical power and polarization state of the signal and pump before they reach that arm's SLM (Santec SLM-100 phase modulator, $1440\times1050$ pixels with a pixel pitch of 10.4 $\mu$m). After each SLM, a second HWP rotates the polarization to the one the crystal accepts. The signal illuminates its SLM with a Gaussian of 2.83 mm $1/e^2$ radius and the pump with 4.0 mm.

The two arms share a 4f system of two lenses of equal focal length, L1 and L2 ($f=200$ mm). The SLMs, combined on a beam splitter, sit at its object plane. A 50 mm MgO-doped PPLN crystal (poling period 19.45 $\mu$m, type-0 phase matching, 40 $^\circ$C) is centered at its Fourier plane, where the focused pump and signal generate sum-frequency output at 775.95 nm. The crystal accepts a band of about 60 GHz in the sum frequency~\cite{Gayer2008}. The pump channel (about 54 GHz) is matched to it, while the signal channel (about 147 GHz) is wider (Sec.~\ref{subsec:limitations}). A short-pass and a narrow band-pass filter follow the crystal. The fiber coupler sits at the image plane after L2. Its asphere and L2 image the mode of a single-mode fiber (SM600) into the crystal with a waist of about 45 $\mu$m, the $\phi_f$ of Eq.~\eqref{eq:2}. Because only light in this one spatial mode reaches the detector, a click measures the coherent projection $|\langle K_j|\psi_x\rangle|^2$ rather than an integrated intensity. An ID Quantique ID100 silicon avalanche photodiode (detection efficiency: around 10\,\% at 776 nm; dark-count rate: as low as 2 counts/s; timing resolution: approximately 40 ps FWHM; dead time: 45 ns) registers the clicks and a Swabian Time Tagger Ultra counts them.

\subsection{Encoding, training and testing}
\label{subsec:encoding-training-and-testing}

The signal SLM encodes a $28\times28$ MNIST digit~\cite{Lecun1998} resized to $784\times784$ pixels, as the phase $\pi v_x$ of Eq.~\eqref{eq:1}. The SLM's phase stroke is short of $2\pi$, so its levels are rescaled in software to give a true $\pi$. Outside the digit the phase is zero, and $A(\bm r)$ is the Gaussian illumination.

The pump SLM writes a complex superposition of $\mathrm{HG}_{mn}$ modes (waist 2.0 mm) by double-phase encoding~\cite{Hsueh1978,Davis1999,MendozaYero2014}. The panel does not render level pairs linearly, so a measured lookup table maps each target to the level pair that produces it (Appendix~\ref{app:d}). The superposition kernels use the ten modes of order $m+n\le3$: $\mathrm{HG}_{00}$, $\mathrm{HG}_{10}$, $\mathrm{HG}_{01}$, $\mathrm{HG}_{20}$, $\mathrm{HG}_{11}$, $\mathrm{HG}_{02}$, $\mathrm{HG}_{30}$, $\mathrm{HG}_{21}$, $\mathrm{HG}_{12}$, $\mathrm{HG}_{03}$.

Each pair of masks (image, kernel) is then measured in its own acquisition of 0.1 s after a fixed SLM settling time. The time tagger records the click times. A calibration loop compensates for polarization drift between the acquisitions (Appendix~\ref{app:d}).

Over the test images the measured masks gave 0.02--720 counts/ms, that is $2\times10^{-7}$ to $7\times10^{-3}$ detected photons per pulse. Two detected photons would fall in one pulse with probability below $3\times10^{-5}$, and the detector registers at most one click per pulse. Each click is therefore assumed to be one converted photon.

The kernels are trained in situ, in three stages, on training digits disjoint from the test digits (Appendix~\ref{app:b}).

\begin{enumerate}
\item \emph{Tomography.} For 40 training digits per class, each digit's $10\times10$ response matrix $gg^\dagger$ is measured, where $g$ collects the projections $g_{mn}(x)$ of Eq.~\eqref{eq:4}. The measurement uses 190 masks: the ten HG modes, and each of the 45 pairs of modes superposed at four relative phases ($0$, $\pi/2$, $\pi$, $3\pi/2$), which separate the real and imaginary parts of the cross terms $g_ag_b^{*}$.
\item \emph{Candidates.} The superposition coefficients $c^{(j)}_{mn}$ are designed on these matrices to maximize the first-click accuracy. A library of 48 kernels is prepared with 19 designed superpositions at two different global phases, along with the ten single basis modes.
\item \emph{Selection.} The library is measured on a further 50 training digits per class. One kernel per class is then chosen from the measured rates by cross-validation, so no prediction of the setup's response enters the final choice. The kernels of the three classes use the same set of basis modes and differ only in their superposition coefficients.
\end{enumerate}

The trained NLONN is tested on a batch of 450 images, 150 per class from the MNIST test set, disjoint from every training digit. For each image, the three selected kernels are written in turn on the pump SLM, and every click is recorded with its arrival time. The three acquisitions are combined by time-division multiplexing. Their clocks are aligned at the start, so the merged record has the same first-click statistics as a pump that cycles through the three kernels faster than photons arrive. The first click in the merged record names the class of its kernel.

\section{Results}
\label{sec:results}

\subsection{Accuracy from first detected photons}
\label{subsec:accuracy-from-recorded-photons}

On the three-class task (digits 2, 4 and 7), the strictest test decides each of the 450 test images by the first photon in its merged record (Sec.~\ref{subsec:encoding-training-and-testing}), a single-shot measurement, which gives 56.2 $\pm$ 2.3\,\%. The merged record holds many clicks, so each image also allows many single-shot trials. The record is divided into trials of 100 $\mu$s, and the first click in each trial names a class. Each image gives a median of 643 trials. Averaged over these repeated trials the accuracy is 53.9 $\pm$ 1.4\,\%. Both estimate the first-click probability of Eq.~\eqref{eq:5}. The repeated trials have the smaller statistical error, and the two agree within 1$\sigma$. Each $\pm$ is the standard error over the 450 test images. On its 150 training digits the selected set reaches 54.9 $\pm$ 2.4\,\%, computed from their measured rates, and the same under five-fold cross-validation, so the choice of kernels does not overfit.

Combining successive trials lets several photons decide together. The decision is naive Bayes for multiple photons: starting from equal priors, each click multiplies the probability of every class by the mean share of that class's training digits at the kernel that clicked, and after the last click the class with the largest posterior is named. Ten photons give 69.7 $\pm$ 1.4\,\%, and Appendix~\ref{app:c} follows the accuracy up to 100 photons and compares the NLONN with other experiments. Since the NLONN is trained to decide from the first photon, Sec.~\ref{subsec:comparison-with-linear-optics} compares it with linear optics at one detected photon.

\begin{figure}[t]
\centering
\includegraphics[width=\columnwidth]{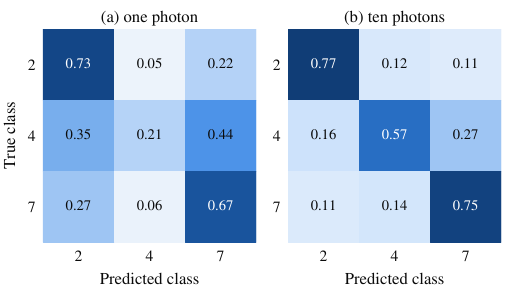}
\caption{Confusion matrices on the 450 test images: the probability that each class is predicted (columns), for test images of each true class (rows), (a) from one detected photon in repeated single-shot trials and (b) from ten detected photons decided together (Appendix~\ref{app:c}). Standard errors are at most 0.03.}
\label{fig:3}
\end{figure}

Figure~\ref{fig:3} shows the NLONN by class. From one photon, twos and sevens are named correctly with probability 0.73 and 0.67, fours with 0.21. The kernel naming class 4 is weak but selective. It clicks least on every digit, so it seldom gives the first click. When it does click, the image is three to four times more likely to be a four than either other digit. With ten photons this selectivity accumulates, and fours are named correctly with probability 0.57, twos and sevens with 0.77 and 0.75.

\subsection{Comparison with linear optics}
\label{subsec:comparison-with-linear-optics}

\begin{figure}[t]
\centering
\includegraphics[width=\columnwidth]{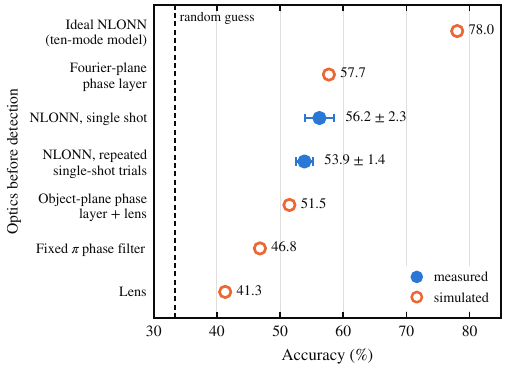}
\caption{Accuracy of the NLONN compared to single-photon detection of the same 450 test images after linear optics. Filled blue symbols are measured in the experiment, with 1$\sigma$ error bars; hollow orange symbols are simulated for the same test images, except the ideal NLONN, which is scored on all 3,042 test images of the three classes. The photon is detected at the Fourier plane of the lens for ``Lens'' and ``Object-plane phase layer + lens'', and at the image plane for every other entry. The fixed filter is a phase plate at the Fourier plane that shifts the central (zero-order) spot by $\pi$. The phase layers are trained in simulation. The ideal NLONN is the best first-click accuracy of kernels drawn from the ten modes, with ideal pump rendering and coherent conversion. Dashed line: random guess, the accuracy of any detection without self-interference for phase-encoded digits.}
\label{fig:4}
\end{figure}

Figure~\ref{fig:4} compares the NLONN with single-photon detection of the same test images after linear optics. Every comparison acts on the same photon states, Eq.~\eqref{eq:1}, and decides from one detected photon. Following Ref.~\cite{Fischbacher2021}, a measurement is classical when it does not let the photon interfere with itself. Its accuracy cannot exceed the best per-pixel lookup table (Appendix~\ref{app:a}), and for phase-encoded digits it is a random guess, $1/C$ for $C$ classes. A unitary $U$ applied through linear optics before a camera lets the photon interfere with itself, with $P(k|x)=|(U\psi_x)(k)|^2$. We compare a lens, $U=\mathcal F$, with the camera at its Fourier plane; fixed phase filters at the Fourier plane, such as a Zernike phase plate~\cite{Zernike1942,Goodman2005}; an object-plane phase layer trained before the lens, $U=\mathcal F\,e^{i\theta}$, equivalent to the one-layer trained-unitary classifier of Ref.~\cite{Fischbacher2021}; and a Fourier-plane phase layer trained in a 4f system, $U=\mathcal F^{-1}e^{i\phi}\mathcal F$, which is a trained convolution~\cite{Goodman2005}. Each is read by a camera of $164\times164$ pixels, which keeps the photon's position. The NLONN is read in one spatial mode by a single-pixel detector, with its kernels shown in turn. The simulations use the signal arm's geometry: the digit, the illumination and the Fourier-plane scale. Each classifier's rule is taken from training images, and every margin below is paired over the same 450 test images.

Detection without self-interference is a random guess (33.3\,\%) for phase-encoded digits. A lens lets the photon interfere with itself, and a camera at its Fourier plane reaches 41.3\,\%. The NLONN exceeds it by 14.9 $\pm$ 2.2 percentage points (pp) in a single shot and by 12.6 $\pm$ 1.2 pp over repeated trials. 

Fixed optics can do better than a lens alone. We scanned seventeen fixed phase filters on a coarse grid (Appendix~\ref{app:a}). The best of them, chosen on training images, places a $\pi$ phase on the zero order at the Fourier plane and reaches 46.8\,\%. The NLONN exceeds it by 9.4 $\pm$ 2.3 pp in a single shot and by 7.1 $\pm$ 1.3 pp over repeated trials.

Trained optics go further. A phase layer at the object plane has 6,724 free phases and reaches 51.5\,\% (Appendix~\ref{app:a}). The NLONN, with 54 trained parameters, exceeds it by 4.7 $\pm$ 2.3 pp in a single shot and by 2.4 $\pm$ 1.4 pp over repeated trials. 

A phase layer at the Fourier plane, a convolution with 26,896 free phases on the $164\times164$ Fourier grid, reaches 57.7\,\%. It is the strongest linear optic we compared. The NLONN is level with it in a single shot ($-$1.5 $\pm$ 2.4 pp) and below it over repeated trials ($-$3.9 $\pm$ 1.4 pp).

The ideal NLONN is a simulation of the same NLONN without the imperfections of the setup. The pump superpositions are rendered exactly and the conversion is fully coherent. Each mode's kernel is computed from Eq.~\eqref{eq:2} for the focused crystal with the calculated fiber mode and the phase mismatch at which HG$_{00}$ converts best, and the kernels are trained on 1,000 training digits per class (Appendix~\ref{app:b}). With the same ten modes it reaches 78.0\,\%. Unlike the linear-optics lines, this number depends on that model.

The NLONN loses the photons its kernels do not convert, and a camera can lose photons too. Behind an absorbing mask that keeps only the purest pixels, a camera reaches 50.0\,\% on the same images, and behind a trained mask 53.8\,\% (Appendix~\ref{app:a}). The NLONN exceeds the first by 6.2 $\pm$ 2.6 pp and is level with the second in a single shot (2.4 $\pm$ 2.6 pp).

\section{Discussion}
\label{sec:discussion}

\subsection{Accuracy and the order of the pump modes}
\label{subsec:accuracy-and-the-order-of-the-pump-modes}

\begin{figure}[t]
\centering
\includegraphics[width=\columnwidth]{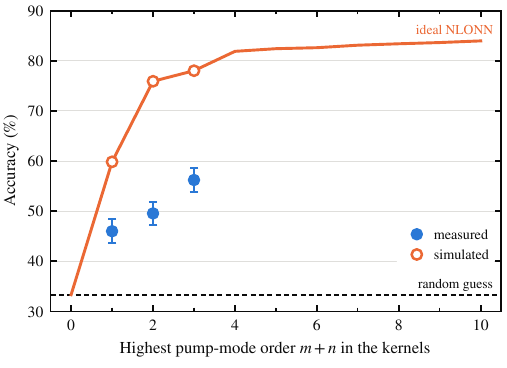}
\caption{Accuracy against the highest pump-mode order $m+n$ in the kernels, up to order 10. Filled blue symbols are measured in the experiment, single shot, with 1$\sigma$ error bars, on the same 450 test images; hollow orange symbols and the orange line are simulated on all 3,042 test images of the three classes. Orange line: the ideal NLONN with one kernel per class drawn from all modes up to that order, including each mode's conversion efficiency. Dashed line: random guess.}
\label{fig:5}
\end{figure}

The kernels of Sec.~\ref{sec:results} are drawn from the ten HG modes of order $m+n\le3$. Figure~\ref{fig:5} shows how the accuracy depends on this choice. Two further runs used the modes of order $\le1$ (three modes) and $\le2$ (six modes). They ran on the same 450 test images and were trained and tested in the same way. In a single shot they reach 46.0 $\pm$ 2.4\,\% and 49.6 $\pm$ 2.4\,\%, against 56.2 $\pm$ 2.3\,\% with ten modes. Over repeated trials they reach 45.2 $\pm$ 1.1\,\% and 48.3 $\pm$ 1.4\,\%, against 53.9 $\pm$ 1.4\,\%. The accuracy rises with every added order. The simulated ideal NLONN follows the same trend: 59.8\,\%, 75.9\,\% and 78.0\,\% for the three mode sets. Beyond that it rises to 84.0\,\% when the kernels draw on all 66 modes of order $\le10$. Higher orders reach finer spatial frequencies of the digits at the Fourier plane, but beyond order 4 each adds less than one point. In the experiment the gain is spread over the orders, and the measured accuracy falls furthest short of the model at order 2.

\subsection{Limitations}
\label{subsec:limitations}

On the three-class task the experiment delivers 0.51 of the ideal NLONN's gain over a random guess, (56.2 $-$ 33.3)/(78.0 $-$ 33.3). The pump beam is slightly aberrated before its SLM and the pump SLM renders level pairs nonlinearly (Appendix~\ref{app:d}). Furthermore, the polarization state of the signal drifts over hours. For this reason kernels are selected from measured rates.

Two limits of the present components account for most of this gap, and neither is a limit of the measurement itself. First, about 11\,\% of each image's response is not coherent. Signal and pump travel at the same group velocity in the crystal, so the phase mismatch depends only on their sum frequency, which the crystal accepts over about 60 GHz~\cite{Gayer2008}. The pump channel matches this acceptance, but the signal channel is wider, and only about a third of the sum frequencies fall inside it (Gaussian estimate). Those outside hardly convert, which lowers the rate equally for every image. Those inside convert with slightly different phase mismatches, so each projects the photon onto a slightly different kernel, and the detector adds them incoherently. A narrower signal filter or a shorter crystal removes the incoherent part, at the cost of rate. Second, the double-phase encoding of the pump leaves a leak: a kernel that should be dark still converts at about 0.05\,\% of the rate of a bright kernel (Appendix~\ref{app:d}). An off-axis pump hologram and a corrected pump beam address it.

The same limits keep the NLONN from an advantage on all ten MNIST classes. A run with six modes of order $m+n\le2$ and one kernel per class was tested on 250 images. It reaches 18.0 $\pm$ 2.4\,\% in a single shot and 17.3 $\pm$ 0.7\,\% over repeated trials. That is level with a lens and camera on the same images (17.3\,\%; paired difference 0.7 $\pm$ 2.9 pp in a single shot). The ideal NLONN with the same modes allows 31.5\,\%. Its ten-class kernels rely on near-cancellations between mode components, so that each kernel stays nearly dark for every class but one. Both limits matter most for such kernels.

The three-class result also depends on the choice of task. The class subset was fixed before any of the tests reported here. All 582 subsets of three to five classes were ranked on a ten-class tomography of training digits, measured with four pump modes, by their cross-validated accuracy minus that of a simulated camera behind an absorbing mask (Appendix~\ref{app:a}). Kernels were then designed and cross-validated on the same tomography for the leading subsets, and digits 2, 4 and 7 had the largest margin. This choice favors a subset on which the NLONN does well against a camera, so the accuracy is not representative of three-class tasks in general.

Finally, the input is a weak coherent beam, and each test image is decided from recorded photons. Each kernel is measured in its own acquisition, and the acquisitions are combined by time-division multiplexing (Sec.~\ref{subsec:encoding-training-and-testing}), so the race is assembled across acquisitions rather than run within one. Each decision is therefore made by a single recorded photon, not by a single input photon. A heralded single-photon input would test the latter directly.

\section{Conclusion}
\label{sec:conclusion}

We demonstrate an NLONN that classifies an image based on the first detected photon. Its trainable layer is SFG with a pump shaped as a superposition of HG modes, which projects the signal photon coherently onto a programmable kernel. Trained in situ using measured responses, it classifies 450 test images of three digits with 56.2 $\pm$ 2.3\,\% accuracy in a single shot, and 53.9 $\pm$ 1.4\,\% over repeated single-shot trials, where detection without self-interference is a random guess (33.3\,\%). It exceeds a lens and camera by 14.9 pp, and, with only 54 parameters, a phase layer trained before the lens with 6,724 parameters by 4.7 pp. It is level with a phase layer trained at the Fourier plane. The NLONN reaches this with one spatial mode on a single-pixel detector, while these linear optics are read by a camera. Resolving the upconverted light in several modes or pixels, for example with an array of single-photon avalanche diodes (SPADs), would give more outcomes per photon and is a route to higher accuracy.

The accuracy rises with the order of the pump modes: from 46.0\,\% to 56.2\,\% in the experiment, and to 84.0\,\% in simulation with all modes up to order 10. The present setup realizes 0.51 of the ideal NLONN's gain over a random guess, so improved rendering and higher-order modes can carry it past the Fourier-plane phase layer. The rest is lost mainly to the incoherent part of the response and to the leak of the pump encoding.

Multi-plane light conversion~\cite{Labroille2014,Fontaine2019,Zhang2023} or an off-axis complex hologram~\cite{Arrizon2007} would render higher-order pump modes with less leak, and a narrower signal band would reduce the incoherent part. A heralded single-photon source~\cite{Ma2020,PhysRevLett.114.110504,Zhang:24,khan2026shaping} would allow testing the NLONN in the quantum-correlated regime, where shaping the pump of a down-conversion source has already been used to train photon correlations for computer vision at low light~\cite{Sohoni2026}.

Because each decision comes from one detected photon, an NLONN suits photon-starved tasks, such as classifying unstained or light-sensitive biological samples and faint scenes in remote sensing~\cite{Zhang2020,Darji2024}. Detected optical energies in such classifiers reach well below a femtojoule per inference, orders of magnitude below digital multiplications~\cite{Wang2022,Ma2025,Anderson2024} (Appendix~\ref{app:c}). The upconverted photon leaves the NLONN in a single mode with its quantum state~\cite{Brecht2015} preserved. The NLONN can therefore also serve as an energy-efficient spatial-mode sorter at the front end of quantum communication~\cite{Erhard2018}.

\begin{acknowledgments}
This research was supported in part by ACC-New Jersey under grant number W15QKN18D0040.
\end{acknowledgments}

\section*{Data availability}
The data and code that support the findings of this article are openly available~\cite{NLONNrepo}.

\appendix
\section{Linear optics and cameras}
\label{app:a}

We first bound any detector that records only which pixel the photon reaches. Let a photon from image $x$ arrive at detector pixel $k$ with probability $P(k|x)$, and let the images of class $c$ have prior $p_c$. The joint distribution is $P(c,k)=p_c\sum_{x\in c}p(x|c)\,P(k|x)$, with $p(x|c)=1/N_c$ for the $N_c$ images of class $c$. Any decision rule, deterministic or stochastic, $q(\hat c|k)$, has accuracy

\begin{equation}
\sum_k\sum_c P(c,k)\,q(c|k) \le \sum_k \max_c P(c,k) ,
\tag{A1}
\label{eq:a1}
\end{equation}

with equality for the per-pixel maximum a posteriori rule. One photon carries one pixel index, so no processing of it, however elaborate, does better~\cite{Fischbacher2021}. Evaluated on the raw MNIST test set, with $P(k|x)$ proportional to brightness, Eq.~\eqref{eq:a1} gives 22.96\,\% for ten classes, the published value, and 54.4\,\% for digits 2, 4 and 7.

For phase digits, a detector without self-interference sees $\sum_{\bm r}M(k|\bm r)|\psi_x(\bm r)|^2$ for some stochastic map $M$. For the states of Eq.~\eqref{eq:1} this does not depend on $x$, so $P(c,k)=p_c\,P(k)$ and the bound is $\max_c p_c$: a random guess.

With a lens, a camera placed at the Fourier plane sees a photon with probability distribution $P(k|x)=|\tilde\psi_x(k)|^2$, computed with the setup's digit size (8.15 mm), illumination (2.83 mm $1/e^2$ radius) and wavelength. The signal plane is sampled on $82\times82$ pixels of 145.5 $\mu$m over a 12 mm window: the digit fills the central $56\times56$ ($2\times2$ pixels per MNIST pixel), and the margin carries the beam's wings at zero phase. Zero-padded to $164\times164$, the Fourier transform gives camera pixels of 12.9 $\mu$m, twice as fine as the field's bandwidth requires; the illumination's focal spot has a radius of 35 $\mu$m. The decision rule is Eq.~\eqref{eq:a1}'s, but with equal priors and the class sums taken from all training images of the subset. It is scored image by image on the 450 test images. Merging pixels can only lower Eq.~\eqref{eq:a1}, since $\sum_{\mathrm{bins}}\max_c\sum_{k\in{\mathrm{bin}}}P(c,k)\le\sum_k\max_cP(c,k)$. We therefore use the full $164\times164$ grid, which resolves every feature of the field. A $64\times64$ crop with an outer bin and raw class sums gives 40.6\,\% instead of 41.3\,\%. A rule fitted on all 3,042 test images of the subset, which include the scored ones, gives 41.5\,\%. A uniform detection efficiency leaves the accuracy per detected photon unchanged, and dark counts can only lower it, so the ideal camera computed here is the generous comparison.

A fixed phase pattern before the camera is untrained, but like the lens it uses the photon's self-interference. We scanned defocus applied at the signal modulator, $\theta=\beta\pi(r/w)^2$ with $\beta\in\{0.25,0.5,1,2,4\}$ and $w$ the illumination radius, as well as Zernike phase contrast~\cite{Zernike1942} at the Fourier plane. In the latter, a plate of phase $\phi\in\{\pm\pi/2,\pi\}$ covers a disk of radius $\rho\in\{0.5,1,2,4\}$ times the illumination's spot at the Fourier plane, and the camera sits at the image plane. Each filter's rule comes from 3,000 training images per class, and the filter is chosen on the same training images. The best is a $\pi$ plate of radius one spot, 46.8\,\% on the 450 test images, and the best defocus ($\beta=1$) reaches 44.3\,\%. A finer scan could find a better fixed filter. Choosing among filters with training data is, however, already a restricted form of training the optic, which the trained phase layers below complete.

The object-plane phase layer of Sec.~\ref{subsec:comparison-with-linear-optics} adds a phase $\theta(\bm r)$ to the image's phase on an $82\times82$ grid over the illuminated beam, with $2\times2$ phase pixels per MNIST pixel. A grid finer than the digit makes this the stronger comparison: it does at least as well as one phase per MNIST pixel. A full-resolution camera sits at the Fourier plane. Training alternates two steps. First, with the pixel-to-class map $d(k)$ fixed, the phases are updated to maximize

\begin{equation}
J(\theta)=\Big\langle\sum_k\delta_{d(k),y_x}\,P_\theta(k|x)\Big\rangle_x ,
\tag{A2}
\label{eq:a2}
\end{equation}

where $P_\theta(k|x)=|(\mathcal Ff_x)(k)|^2$ is the probability that the photon reaches camera pixel $k$. Here $f_x=\psi_x e^{i\theta}$ is the photon's state of Eq.~\eqref{eq:1} after the phase layer, which adds $\theta$ to the digit's encoded phase, and $\mathcal F$ is the Fourier transform performed by the lens. The state $\psi_x$ is normalized, and neither the phase layer nor the lens changes the total probability (Parseval's theorem), so $\sum_kP_\theta(k|x)=1$. $J$ is then the probability that it lands on a pixel of its own class, averaged over the training images $\langle\cdot\rangle_x$. The Kronecker delta $\delta_{d(k),y_x}$ is 1 when pixel $k$ names the image's class $y_x$ and 0 otherwise; over the classes it is the one-hot form of the map $d$. The map starts as the lens-and-camera rule ($\theta=0$). The phases are updated by Adam~\cite{Kingma2015}, a gradient-ascent method, which needs the gradient of $J$ with respect to every phase pixel. We supply it analytically. For one image,

\begin{equation}
\frac{\partial J_x}{\partial\theta}=-2\,\mathrm{Im}\!\left[f_x\,\overline{\mathcal F^{-1}\!\left(W_x\,\mathcal Ff_x\right)}\right] .
\tag{A3}
\label{eq:a3}
\end{equation}

\begin{figure}[t]
\centering
\includegraphics[width=\columnwidth]{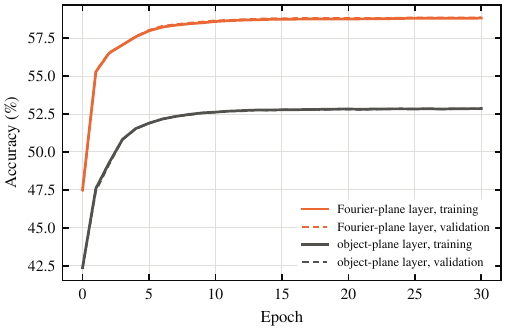}
\caption{Training of the two phase layers in simulation: accuracy per epoch on the 16,258 training images (solid) and the 1,807 validation images (dashed) of digits 2, 4 and 7. For each layer the training and validation curves agree within 0.1 pp.}
\label{fig:6}
\end{figure}

The mask $W_x(k)=\delta_{d(k),y_x}$ keeps the camera pixels of the image's own class, and the bar denotes the complex conjugate. Read from the inside out, the field is propagated to the camera, the pixels of the other classes are removed, the remainder is propagated back to the phase layer, and its overlap with the field there gives the gradient at each phase pixel. This costs one forward and one inverse Fourier transform per image, like back-propagation through a linear layer. Equation~\eqref{eq:a3} agrees with finite differences to $10^{-6}$. Second, $d(k)$ is refitted as the equal-prior maximum a posteriori rule. Training on 16,258 images of digits 2, 4 and 7 levels off after about 15 epochs (Fig.~\ref{fig:6}). The epoch kept is the best on 1,807 validation images. A simulation gives the full probability distribution of where the photon lands. Each test image is therefore scored by the probability that its photon lands on a pixel of the right class. A single-shot measurement would classify the image correctly with this same probability. On the 450 test images it reaches 51.5\,\%. At $\theta=0$ the procedure reproduces the lens and camera.

The Fourier-plane phase layer of Sec.~\ref{subsec:comparison-with-linear-optics} acts on the photon's state: $\psi_x$ is Fourier transformed, multiplied by $e^{i\phi(\bm u)}$ on the $164\times164$ Fourier grid, and transformed back; the camera at the output plane records $|g_x(k)|^2$ with $g_x=\mathcal F^{-1}[e^{i\phi}\mathcal F\psi_x]$. Training alternates as for the object-plane layer, with the gradient through two FFTs checked against finite differences, and starts from the best fixed filter (the $\pi$ plate). It is trained on the same 16,258 images, and the epoch kept is the best on the same 1,807 validation images (epoch 30 of 30; Fig.~\ref{fig:6}). It reaches 57.7\,\% on the 450 test images and 59.0\,\% on all 3,042 test images of the subset. Like a Zernike phase plate, a Fourier-plane phase turns the digit's phase into intensity at each image pixel, so its camera sees the digit's shape.

A classical camera can lose photons too. Behind an absorbing mask it decides from the first photon that gets through, the same protocol as the NLONN's but without self-interference after the lens. A measurement allowed such an inconclusive outcome can do better on its conclusive results~\cite{Barnett2009,Eldar2003}.

A mask before the Fourier-plane camera passes a photon that lands on pixel $k$ with probability $w_k$ and absorbs it otherwise. The camera decides from the first photon that passes, by the class $d(k)$ its pixel names. Its accuracy is the probability that a photon that passes lands on a pixel of the right class, $\sum_kw_kP(k|x)\,\delta_{d(k),y_x}/p_x$, averaged over test images, with $p_x=\sum_kw_kP(k|x)$ the probability that the photon passes.

We tried two masks. The purity-ranked mask keeps whole pixels, the purest first, where purity is the share of training photons from the most common class; a pixel without class information has purity 0.33. The 86 purest pixels, of purity 0.53--0.58, pass 1.3\,\% of the photons; the brightest pixel, at the center of the focal spot, has purity only 0.38. The trained mask starts from these pixels and refines its transmissions on training images; it keeps ten pixels and passes about 0.01\,\% of the photons. On the 450 test images the purity-ranked mask reaches 50.0\,\% and the trained mask 53.8\,\%. Both are achievable classical lines, not bounds: a better mask may exist.
\section{Training the NLONN}
\label{app:b}

\emph{Objective.} For kernels $j$ with weights $t_j$ and rates $R_j(x)$, the first click comes from kernel $j$ with probability $P_t(j|x)=t_jR_j(x)/\sum_{j'}t_{j'}R_{j'}(x)$. The rates are predicted from the tomography when candidates are designed, and measured when kernels are selected. The weights act as dwell times. They are free in the design, and the selected NLONN uses equal ones. Let $L_{jc}$ be the mean of $P_t(j|x)$ over the training digits of class $c$. Each kernel names the class with the largest $L_{jc}$, so its decision is $D_{jc}=1$ for that class and 0 for the others. The accuracy is the mean probability that the first click names the true class $y_x$,

\begin{equation}
A=\frac1N\sum_x\sum_jP_t(j|x)\,D_{j\,y_x} ,
\tag{B1}
\label{eq:b1}
\end{equation}

with the sum over the $N$ training digits. The hard decision changes in steps, so $A$ gives no useful gradient. For optimization we replace $D_{jc}$ by a softmax over classes,

\begin{equation}
\tilde D_{jc}=\frac{\exp\big(\beta L_{jc}/L_j^{\max}\big)}{\sum_{c'}\exp\big(\beta L_{jc'}/L_j^{\max}\big)} ,\qquad L_j^{\max}=\max_cL_{jc} ,
\tag{B2}
\label{eq:b2}
\end{equation}

with $\beta$ an inverse temperature. A class whose mean lies a fraction $\delta$ below the kernel's best receives the weight $e^{-\beta\delta}$ relative to it. A small $\beta$ spreads each kernel's vote over nearly tied classes and gives smooth gradients. As $\beta\to\infty$, $\tilde D$ becomes the hard decision. We optimize at $\beta=30$ and then refine at $\beta=100$. Reported accuracies always use the hard decision, Eq.~\eqref{eq:b1}.

\begin{figure}[t]
\centering
\includegraphics[width=\columnwidth]{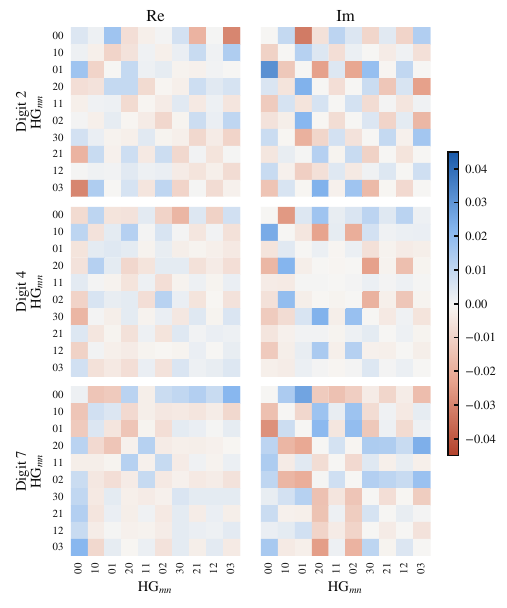}
\caption{Measured response of each class. Each panel shows $\langle M_x\rangle_c-\langle M_x\rangle$, the mean trace-normalized response matrix of the 40 training digits of class $c$ minus the mean over all three classes, in the basis of the ten pump modes $\mathrm{HG}_{mn}$. Left: real part; right: imaginary part. The classes differ most in the off-diagonal elements.}
\label{fig:7}
\end{figure}

\emph{Tomography.} The kernels are superpositions of the ten modes of order $m+n\le3$ of Sec.~\ref{subsec:encoding-training-and-testing}. For the tomography we number them $a=1,\dots,10$: $g_a(x)$ is the projection $g_{mn}(x)$ of Eq.~\eqref{eq:4} for mode $a$, and $c_a$ is its coefficient $c_{mn}$. Equation~\eqref{eq:4} then reads $p_j(x)=\eta\,c^{(j)\mathsf T}M_x\,c^{(j)*}$, with the $10\times10$ response matrix $(M_x)_{ab}=g_a(x)\,g_b^*(x)$. Mode $a$ alone gives a diagonal element, $p\propto(M_x)_{aa}=|g_a(x)|^2$. A pair of modes $a<b$ is shown as the mask $c=(e_a+e^{i\varphi}e_b)/s_{ab}$. Here $e_a$ is the unit vector of mode $a$, and $s_{ab}$ scales the superposed field to the modulator's unit amplitude. By Eq.~\eqref{eq:4} the pair's rate is

\begin{equation}
\begin{split}R^{\varphi}_{ab}&\propto\frac{|g_a+e^{i\varphi}g_b|^2}{s_{ab}^2}=A+B\cos\varphi+C\sin\varphi ,\\A&=\frac{(M_x)_{aa}+(M_x)_{bb}}{s_{ab}^2} ,\\B&=\frac{2\,\mathrm{Re}(M_x)_{ab}}{s_{ab}^2} ,\quad C=\frac{2\,\mathrm{Im}(M_x)_{ab}}{s_{ab}^2} .\end{split}
\tag{B3}
\label{eq:b3}
\end{equation}

\begin{figure}[t]
\centering
\includegraphics[width=\columnwidth]{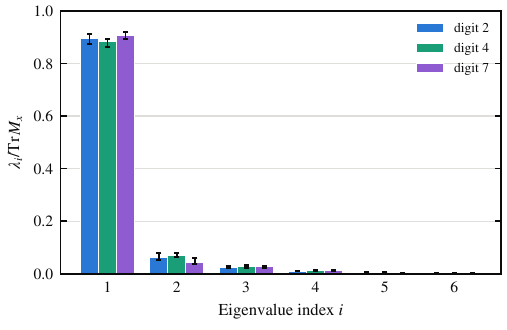}
\caption{Eigenvalue spectra of the measured response matrices. Bars: median of the $i$-th eigenvalue of $M_x/\mathrm{Tr}\,M_x$ over the 40 training digits of each class; error bars: interquartile range. A fully coherent projection has a single eigenvalue of 1. The first eigenvalue holds 0.89 of the response; the rest are spread over a few weak components.}
\label{fig:8}
\end{figure}

This is the fringe of the two modes' conversion amplitudes as their relative phase $\varphi$ is stepped. The four phases $0$, $\pi/2$, $\pi$ and $3\pi/2$ give $B=(R^{0}_{ab}-R^{\pi}_{ab})/2$ and $C=(R^{\pi/2}_{ab}-R^{3\pi/2}_{ab})/2$, and so $(M_x)_{ab}=s_{ab}^2(B+iC)/2$. The offset $A$ only repeats the singles, together with any background, and is not used; the differences cancel the background. $M_x$ is measured this way for each of 120 training digits (40 per class), with 190 masks: the ten single modes and the 45 pairs at four phases. $M_x$ is Hermitian, with 100 real parameters, and the 190 rates over-determine them. The fit is projected onto positive-semidefinite matrices. Figure~\ref{fig:7} shows how the classes differ. If the projection is fully coherent, $M_x=g\,g^\dagger$ has a single nonzero eigenvalue. Incoherent conversion adds further eigenvalues. We measure coherence by the largest eigenvalue $\lambda_{\max}$ of $M_x$ divided by the sum of all its eigenvalues, $\mathrm{Tr}\,M_x$. This ratio is 1 for a fully coherent projection; in the experiment its median is 0.894 (interquartile range 0.873--0.908; Fig.~\ref{fig:8}). The incoherent part comes from the signal bandwidth (Sec.~\ref{subsec:limitations}).

\begin{figure*}[t]
\centering
\includegraphics[width=\textwidth]{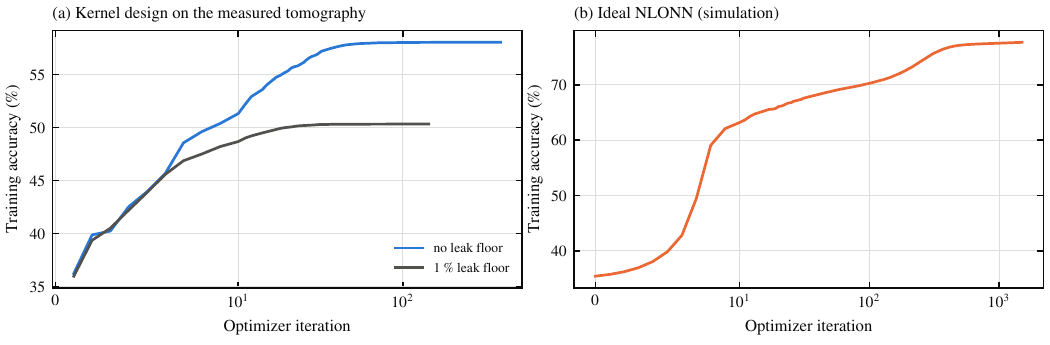}
\caption{Training of the NLONN's kernels. (a) Candidate design on the measured tomography of the 120 training digits: training accuracy predicted from the tomography after each iteration of the quasi-Newton optimizer. Blue: no leak floor. Dark: a leak floor that adds 1\,\% of each digit's summed single-mode rate to the predicted rate of every kernel (about 2\,\% of the HG$_{00}$ rate). The floor stands in for the counts the setup adds to every kernel, so the design cannot separate the classes by driving kernels nearly dark. (b) The ideal NLONN with the ten modes, trained by gradient ascent on 1,000 training digits per class. Its objective is the training accuracy; its test accuracy follows within about 1 pp and ends at 78.0\,\%.}
\label{fig:9}
\end{figure*}

\emph{Candidates.} Candidate kernels are designed on the measured $M_x$, which predicts the rate of kernel $j$ as $R_j(x)=c^{(j)\mathsf T}M_xc^{(j)*}$. Each $M_x$ is first divided by its trace, the summed rate of the ten single modes, so that bright and dim digits count equally. A design is a set of ten kernels with weights. Their coefficients, each $c^{(j)}$ kept at unit norm, and the weights maximize the accuracy of Eq.~\eqref{eq:b1} with the soft decision of Eq.~\eqref{eq:b2}. The optimization is started from three kinds of initial kernels, and the best result is kept: the ten single modes; the rank-one parts of the minimum-error measurement~\cite{Helstrom1976} that discriminates the class means of $M_x$; and random superpositions. Before the modulator calibration of Appendix~\ref{app:d}, a rendered kernel's rate carried an error of about 0.01--0.02 of the HG$_{00}$ rate that changed from mask to mask. A design that separates the classes by cancelling to near-zero rates therefore fails in the experiment. Adding a floor $f$ to the diagonal of every normalized $M_x$ prevents this: every kernel then predicts a rate of at least $f$. On this scale HG$_{00}$ alone gives about 0.57, so $f=0.01$ is about 2\,\% of its rate. Two designs are made, at $f=0$ and $f=0.01$. They predict 58.1\,\% and 50.3\,\% on the training digits (Fig.~\ref{fig:9}a). Their 20 kernels and ten rank-one parts of the minimum-error measurement leave 19 distinct superpositions once near-duplicates are removed. Each is written at two global phases (90$^\circ$ and 270$^\circ$). With the ten single modes they form the library of 48 masks.

\emph{Selection.} The tomography predicts the designed kernels' rates only roughly: relative to HG$_{00}$, their measured class-mean rates are 0.21--1.35 of the predicted ones (median 0.80). Kernels are therefore selected from measured rates. The library of 48 masks is measured on 150 fresh training digits (50 per class). Its rates are corrected for the detector's dead time (50 ns as set on the time tagger, at most 4\,\% at the highest rates) and the dark rate (2 counts/s), and normalized to the reference kernel (Appendix~\ref{app:d}). One kernel per class is then chosen at equal dwells. Each class's slot in turn takes the library mask that most raises the training accuracy, Eq.~\eqref{eq:b1} with equal weights, starting from the mask with the largest lead for that class. Nested five-fold cross-validation, which repeats the choice within each training fold, estimates its accuracy. The test measured 16 of the 48 masks, and the choice is the same whether made among those or among all 48. The runs of Sec.~\ref{subsec:accuracy-and-the-order-of-the-pump-modes} are trained the same way.

\emph{Model of the ideal NLONN.} The model of Figs.~\ref{fig:4} and~\ref{fig:5} computes each mode's kernel $\kappa_{mn}$ from Eq.~\eqref{eq:2} for the focused crystal with the calculated fiber mode, in the signal arm's geometry. Inside the crystal the beams spread according to its refractive index~\cite{Gayer2008}. The phase mismatch is the one at which HG$_{00}$ converts a blank signal best, $\Delta k\approx66$ m$^{-1}$, as the oven temperature is set for maximum conversion. It is not zero because of the Gouy phases of the focused beams. HG$_{00}$ converts at full efficiency and higher orders at their computed ratio to it. For each mode set, one kernel per class is optimized within the span of the set by gradient ascent on the mean first-click probability of Eq.~\eqref{eq:5}, over 1,000 training digits per class (Fig.~\ref{fig:9}b). It is scored on all 3,042 test digits of the three classes. With the ten modes it reaches 78.0\,\% and converts 0.128\,\% of the photons (75.6\,\% on the 450 test images of the experiment).

The kernel naming class 4 is the weakest. Relative to HG$_{00}$ on the same digits, it renders at 0.28--0.43 of the rate its tomography predicts, against 0.54--1.35 for the other two kernels. It is also the dimmest of the three by design. On fours its mean rate is therefore about half that of the kernels naming 2 and 7, so it seldom gives the first click (Fig.~\ref{fig:3}).

\section{A comparative study of photon-based classification}
\label{app:c}

Each repeated single-shot trial of Sec.~\ref{subsec:accuracy-from-recorded-photons} gives an independent photon, so the same measurement shows how the accuracy grows when several photons decide. The photons are combined by the naive Bayes rule given there, applied after $n$ clicks; time slots without a click are ignored, and at $n=1$ it is the first-click decision. The ideal NLONN is scored the same way, from photons drawn from its simulated kernel probabilities, with class shares from its training digits.

\begin{figure}[t]
\centering
\includegraphics[width=\columnwidth]{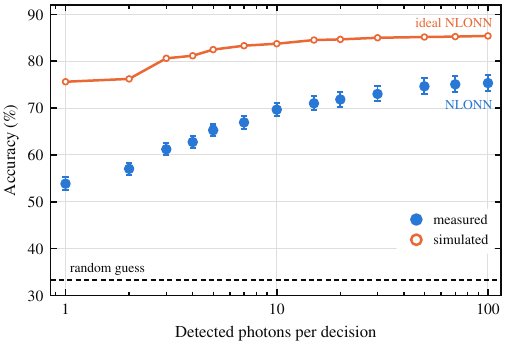}
\caption{Accuracy against the number of detected photons per decision on the 450 test images: the NLONN from its recorded clicks (filled, $\pm$1$\sigma$ over images) and the ideal NLONN from simulated photons (hollow). Dashed line: random guess.}
\label{fig:10}
\end{figure}

On the 450 test images the NLONN rises from 53.9 $\pm$ 1.4\,\% at one photon to 69.7 $\pm$ 1.4\,\% at ten and 75.3 $\pm$ 1.8\,\% at a hundred, and the ideal NLONN from 75.6\,\% to 85.4\,\% (Fig.~\ref{fig:10}). Both gain most within the first ten photons and then level off.

The level is set by the number of outcomes. A photon at the NLONN has three outcomes, one per kernel, so many photons only estimate the image's three kernel probabilities. Classifying even the exact probabilities, by their 15 nearest training digits, gives 84.2\,\% for the ideal NLONN and 78.9\,\% for the measured kernels, close to where the curves level off. The kernels were trained for the first photon, one per class. A measurement trained for one photon is not the one that gains most from many: Duplinskiy et al. trained a diffractive network for each regime and found the first better at one photon and the second better as photons accumulate~\cite{Duplinskiy2026}. Higher accuracy from many photons would call for more kernels than classes, with the decision taken from their counts.

\begin{figure}[t]
\centering
\includegraphics[width=\columnwidth]{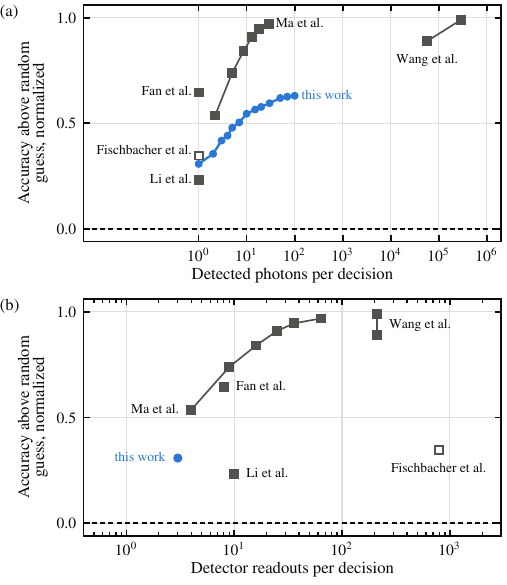}
\caption{Accuracy above the random guess, normalized as $(A-1/C)/(1-1/C)$, for classifiers of MNIST digits, against (a) detected photons per decision and (b) detector readouts per decision, the pixels read times the measurement settings shown in turn. Blue: this work (three classes; one detector, three kernels; from Fig.~\ref{fig:10}). It keeps three readouts for every number of photons, so (b) shows its first photon only; with 100 photons it rises to 0.63 at the same three readouts. Gray, filled: experiments by Li et al.~\cite{Li2024}, Fan et al.~\cite{Fan2026} (eight classes), Ma et al.~\cite{Ma2026} and Wang et al.~\cite{Wang2022} (ten classes). Gray, hollow: the theory by Fischbacher et al. for a single photon with interference~\cite{Fischbacher2021} (ten classes). For Wang et al., photons per decision are its photons per multiplication times its 89,400 multiplications per inference, and its readouts are its 210 dot products.}
\label{fig:11}
\end{figure}

Figure~\ref{fig:11} places the NLONN among other work that classifies MNIST digits at the few-photon level. Because the tasks differ in their number of classes $C$, the accuracy $A$ is shown above the random guess and normalized, $(A-1/C)/(1-1/C)$, which is 0 for a random guess and 1 for a perfect classifier. Fewer classes remain easier, so the comparison is indicative. Figure~\ref{fig:11}(b) counts the detector readouts per decision: the pixels read times the measurement settings shown in turn. For ten classes, no measurement of a single-photon without interference exceeds 22.96\,\%, and a designed unitary before 790 detectors reaches 41.27\,\% in theory~\cite{Fischbacher2021}. Li et al. measured 31.0\,\% from one photon with ten phase masks shown in turn before upconversion into a single mode in a PPLN waveguide~\cite{Li2024}. Fan et al. measured 69.0\,\% on eight classes from one photon with a trained phase layer and eight pixels of a single-photon detector array~\cite{Fan2026}. Optical neural networks near the quantum limit reach 58\,\% on ten classes from about two detected photons, and 90\,\% and above from about 13 to tens of thousands. They also use a single-pixel detector, but collect many spatial modes, take 4 to 210 readouts in turn and combine the counts electronically~\cite{Ma2026,Wang2022}. The NLONN decides from one detected photon among three readouts of one detector, with a trained nonlinear measurement and no electronic processing.

In detected energy, one photon at 776 nm is 0.26 aJ per decision, against about 0.8 aJ for Ref.~\cite{Ma2026} and 20 fJ for Ref.~\cite{Wang2022}. The detected light is not the whole optical cost, because the pump runs while the NLONN waits for a click, about 0.1 ms on the present setup. With the pump power estimated from a calibration of the setup, this amounts to some tens of nanojoules per decision. That is comparable to computing the same three overlaps digitally (about $10^4$ multiply-accumulates at 0.3--20 pJ each~\cite{Hamerly2019,Anderson2024}), which would further need a camera image of the scene. Because the click rate grows with the pump power, the pump energy per decision is set by the conversion per signal photon, not by the pump power. A waveguide~\cite{Li2024} and a collection matched to the converted mode would lower it by orders of magnitude. Like the references above, these estimates count optical energy only, not the power to run the equipment.

\section{Pump modulator calibration and drift}
\label{app:d}

The Santec SLM-100 panels reach 1.56$\pi$ (signal) and 1.86$\pi$ (pump) over their 1,024 levels, not 2$\pi$. The signal levels are rescaled so that a phase of $\pi$ is a measured true $\pi$. The pump levels are chosen through the lookup below.

A complex pump coefficient $ae^{i\varphi}$ is written as two phases $\varphi\pm\arccos a$ on neighboring superpixels of $8\times8$ pixels, arranged as a checkerboard, so that their average field is $ae^{i\varphi}$~\cite{Hsueh1978}. Each phase is set by one of the panel's 1,024 levels. On the panel, the field that the two levels produce is not this average. We measured the field of every pair of 24 levels spread over the stroke (300 pairs). From these we built a lookup table that maps a target amplitude and phase to the pair whose measured field is closest. Over the unit disk the table's field deviates from the target by a median of 0.005 of the unit amplitude. With the same table, a dark mask converts at $4.6\times10^{-4}$ of the rate of a bright one.

A calibration loop corrects the drift due to the signal polarization during a run. Every 40 acquisitions, calibration masks are shown on both modulators. The signal amplifier is then servoed back to a fixed rate of about 1,800 counts/ms. Every 20 acquisitions a reference kernel ($\mathrm{HG}_{00}$ on the pump against the signal calibration mask) is measured for 10 ms. Every library rate is divided by the references interpolated to its acquisition time, relative to their mean. The dark rate is measured once per run with both amplifiers off and subtracted.

The reference kernels also flag the drift that the loop leaves. An acquisition is flagged when its interpolated reference lies beyond $\pm$15\,\% of the run's starting level. In the test, 0 of 3,150 acquisitions were flagged with three modes, 57 of 4,950 with six, 13 of 7,200 with ten, and 0 of 4,000 in the ten-class run. Dropping the test images they touch changes the single-shot accuracy by at most 0.2 pp.

\bibliography{references}

@article{Eckstein2011,
  author  = {Eckstein, Andreas and Brecht, Benjamin and Silberhorn, Christine},
  title   = {A quantum pulse gate based on spectrally engineered sum frequency generation},
  journal = {Optics Express},
  volume  = {19},
  number  = {15},
  pages   = {13770--13778},
  year    = {2011},
  doi = {10.1364/oe.19.013770}
}

@misc{khan2026shaping,
  author = {Khan, Usman and Lopez, Daniel and Padmasola, Venkat Aneesh and Dangi, Nirmal and Kumar, Santosh and Huang, Yu-Ping},
  title  = {Shaping frequency-bin biphoton correlations via programmable optical frequency combs},
  year   = {2026},
  eprint = {2611.09713},
  archivePrefix = {arXiv}
}

@Article{Zhou2026ER,
author={Zhou, Qingyi and Kim, Jungmin and Tao, Yutian and Huang, Guoming and Zhou, Ming and Shao, Zewei and Yu, Zongfu},
title={Quantum nonlinearity for optical neural computing},
journal={Nature Communications},
year={2026},
month={Aug},
day={27},
volume={17},
number={1},
pages={10241},
doi={10.1038/s41467-026-77202-y},
}

@article{Zhang:24,
  author  = {Zhang, Zheshen and You, Chenglong and {Maga{\~n}a-Loaiza}, Omar S. and Fickler, Robert and Le\'{o}n-Montiel, Roberto de J. and Torres, Juan P. and Humble, Travis S. and Liu, Shuai and Xia, Yi and Zhuang, Quntao},
  title   = {Entanglement-based quantum information technology: a tutorial},
  journal = {Advances in Optics and Photonics},
  volume  = {16},
  number  = {1},
  pages   = {60--162},
  year    = {2024},
  doi = {10.1364/aop.497143}
}

@article{PhysRevLett.114.110504,
  author  = {Cai, X.-D. and Wu, D. and Su, Z.-E. and Chen, M.-C. and Wang, X.-L. and Li, Li and Liu, N.-L. and Lu, C.-Y. and Pan, J.-W.},
  title   = {Entanglement-Based Machine Learning on a Quantum Computer},
  journal = {Physical Review Letters},
  volume  = {114},
  number  = {11},
  pages   = {110504},
  year    = {2015},
  doi = {10.1103/physrevlett.114.110504}
}

@article{Brecht2015,
  author  = {Brecht, Benjamin and Reddy, Dileep V. and Silberhorn, Christine and Raymer, Michael G.},
  title   = {Photon temporal modes: a complete framework for quantum information science},
  journal = {Physical Review X},
  volume  = {5},
  pages   = {041017},
  year    = {2015},
  doi = {10.1103/physrevx.5.041017}
}

@article{Shahverdi2017,
  author  = {Shahverdi, Amin and Sua, Yong Meng and Tumeh, Lubna and Huang, Yu-Ping},
  title   = {Quantum parametric mode sorting: Beating the time-frequency filtering},
  journal = {Scientific Reports},
  volume  = {7},
  pages   = {6495},
  year    = {2017},
  doi = {10.1038/s41598-017-06564-7}
}

@article{Kumar2019,
  author  = {Kumar, Santosh and Zhang, He and Maruca, Stephen and Huang, Yu-Ping},
  title   = {Mode-selective image upconversion},
  journal = {Optics Letters},
  volume  = {44},
  number  = {1},
  pages   = {98--101},
  year    = {2019},
  doi = {10.1364/ol.44.000098}
}

@inproceedings{Fischbacher2021,
  author    = {Fischbacher, Thomas and Sbaiz, Luciano},
  title     = {Single-photon image classification},
  booktitle = {International Conference on Learning Representations (ICLR)},
  year      = {2021},
  eprint = {2008.05859},
  archivePrefix = {arXiv}
}

@misc{Li2024,
  author = {Li, Lili and Kumar, Santosh and Garikapati, Malvika and Huang, Yu-Ping},
  title  = {First photon machine learning},
  year   = {2024},
  eprint = {2410.17471},
  archivePrefix = {arXiv}
}

@article{Hsueh1978,
  author  = {Hsueh, C. K. and Sawchuk, A. A.},
  title   = {Computer-generated double-phase holograms},
  journal = {Applied Optics},
  volume  = {17},
  number  = {24},
  pages   = {3874--3883},
  year    = {1978},
  doi = {10.1364/ao.17.003874}
}

@article{Davis1999,
  author  = {Davis, Jeffrey A. and Cottrell, Don M. and Campos, Juan and Yzuel, Mar\'ia J. and Moreno, Ignacio},
  title   = {Encoding amplitude information onto phase-only filters},
  journal = {Applied Optics},
  volume  = {38},
  number  = {23},
  pages   = {5004--5013},
  year    = {1999},
  doi = {10.1364/ao.38.005004}
}

@article{MendozaYero2014,
  author  = {Mendoza-Yero, Omel and M\'inguez-Vega, Gladys and Lancis, Jesus},
  title   = {Encoding complex fields by using a phase-only optical element},
  journal = {Optics Letters},
  volume  = {39},
  number  = {7},
  pages   = {1740--1743},
  year    = {2014},
  doi = {10.1364/ol.39.001740}
}

@book{Goodman2005,
  author    = {Goodman, Joseph W.},
  title     = {Introduction to Fourier Optics},
  edition   = {3rd},
  publisher = {Roberts \& Company},
  address   = {Englewood, CO},
  year      = {2005}
}

@article{Shen2017,
  author  = {Shen, Yichen and Harris, Nicholas C. and Skirlo, Scott and Prabhu, Mihika and Baehr-Jones, Tom and Hochberg, Michael and Sun, Xin and Zhao, Shijie and Larochelle, Hugo and Englund, Dirk and Soljačić, Marin},
  title   = {Deep learning with coherent nanophotonic circuits},
  journal = {Nature Photonics},
  volume  = {11},
  number  = {7},
  pages   = {441--446},
  year    = {2017},
  doi = {10.1038/nphoton.2017.93}
}

@article{Lin2018,
  author  = {Lin, Xing and Rivenson, Yair and Yardimci, Nezih T. and Veli, Muhammed and Luo, Yi and Jarrahi, Mona and Ozcan, Aydogan},
  title   = {All-optical machine learning using diffractive deep neural networks},
  journal = {Science},
  volume  = {361},
  number  = {6406},
  pages   = {1004--1008},
  year    = {2018},
  doi = {10.1126/science.aat8084}
}

@article{Wetzstein2020,
  author  = {Wetzstein, Gordon and Ozcan, Aydogan and Gigan, Sylvain and Fan, Shanhui and Englund, Dirk and Soljačić, Marin and Denz, Cornelia and Miller, David A. B. and Psaltis, Demetri},
  title   = {Inference in artificial intelligence with deep optics and photonics},
  journal = {Nature},
  volume  = {588},
  number  = {7836},
  pages   = {39--47},
  year    = {2020},
  doi = {10.1038/s41586-020-2973-6}
}

@article{McMahon2023,
  author  = {McMahon, Peter L.},
  title   = {The physics of optical computing},
  journal = {Nature Reviews Physics},
  volume  = {5},
  number  = {12},
  pages   = {717--734},
  year    = {2023},
  doi = {10.1038/s42254-023-00645-5}
}

@article{Wang2022,
  author  = {Wang, Tianyu and Ma, Shi-Yuan and Wright, Logan G. and Onodera, Tatsuhiro and Richard, Brian C. and McMahon, Peter L.},
  title   = {An optical neural network using less than 1 photon per multiplication},
  journal = {Nature Communications},
  volume  = {13},
  number  = {1},
  pages   = {123},
  year    = {2022},
  doi = {10.1038/s41467-021-27774-8}
}

@article{Shastri2021,
  author  = {Shastri, Bhavin J. and Tait, Alexander N. and Ferreira de Lima, T. and Pernice, Wolfram H. P. and Bhaskaran, Harish and Wright, C. D. and Prucnal, Paul R.},
  title   = {Photonics for artificial intelligence and neuromorphic computing},
  journal = {Nature Photonics},
  volume  = {15},
  number  = {2},
  pages   = {102--114},
  year    = {2021},
  doi = {10.1038/s41566-020-00754-y}
}

@article{Feldmann2021,
  author  = {Feldmann, J. and Youngblood, N. and Karpov, M. and Gehring, H. and Li, X. and Stappers, M. and Le Gallo, M. and Fu, X. and Lukashchuk, A. and Raja, A. S. and Liu, J. and Wright, C. D. and Sebastian, A. and Kippenberg, T. J. and Pernice, W. H. P. and Bhaskaran, H.},
  title   = {Parallel convolutional processing using an integrated photonic tensor core},
  journal = {Nature},
  volume  = {589},
  number  = {7840},
  pages   = {52--58},
  year    = {2021},
  doi = {10.1038/s41586-020-03070-1}
}

@article{Xu2021,
  author  = {Xu, Xingyuan and Tan, Mengxi and Corcoran, Bill and Wu, Jiayang and Boes, Andreas and Nguyen, Thach G. and Chu, Sai T. and Little, Brent E. and Hicks, Damien G. and Morandotti, Roberto and Mitchell, Arnan and Moss, David J.},
  title   = {11 {TOPS} photonic convolutional accelerator for optical neural networks},
  journal = {Nature},
  volume  = {589},
  number  = {7840},
  pages   = {44--51},
  year    = {2021},
  doi = {10.1038/s41586-020-03063-0}
}

@article{Zhou2021,
  author  = {Zhou, Tiankuang and Lin, Xing and Wu, Jiamin and Chen, Yitong and Xie, Hao and Li, Yipeng and Fan, Jingtao and Wu, Huaqiang and Fang, Lu and Dai, Qionghai},
  title   = {Large-scale neuromorphic optoelectronic computing with a reconfigurable diffractive processing unit},
  journal = {Nature Photonics},
  volume  = {15},
  number  = {5},
  pages   = {367--373},
  year    = {2021},
  doi = {10.1038/s41566-021-00796-w}
}

@article{Ashtiani2022,
  author  = {Ashtiani, Farshid and Geers, Alexander J. and Aflatouni, Firooz},
  title   = {An on-chip photonic deep neural network for image classification},
  journal = {Nature},
  volume  = {606},
  number  = {7914},
  pages   = {501--506},
  year    = {2022},
  doi = {10.1038/s41586-022-04714-0}
}

@article{Hamerly2019,
  author  = {Hamerly, Ryan and Bernstein, Liane and Sludds, Alexander and Soljačić, Marin and Englund, Dirk},
  title   = {Large-Scale Optical Neural Networks Based on Photoelectric Multiplication},
  journal = {Physical Review X},
  volume  = {9},
  number  = {2},
  pages   = {021032},
  year    = {2019},
  doi = {10.1103/physrevx.9.021032}
}

@article{Zuo2019,
  author  = {Zuo, Ying and Li, Bohan and Zhao, Yujun and Jiang, Yue and Chen, You-Chiuan and Chen, Peng and Jo, Gyu-Boong and Liu, Junwei and Du, Shengwang},
  title   = {All-optical neural network with nonlinear activation functions},
  journal = {Optica},
  volume  = {6},
  number  = {9},
  pages   = {1132},
  year    = {2019},
  doi = {10.1364/optica.6.001132}
}

@article{Sludds2022,
  author  = {Sludds, Alexander and Bandyopadhyay, Saumil and Chen, Zaijun and Zhong, Zhizhen and Cochrane, Jared and Bernstein, Liane and Bunandar, Darius and Dixon, P. Ben and Hamilton, Scott A. and Streshinsky, Matthew and Novack, Ari and Baehr-Jones, Tom and Hochberg, Michael and Ghobadi, Manya and Hamerly, Ryan and Englund, Dirk},
  title   = {Delocalized photonic deep learning on the internet’s edge},
  journal = {Science},
  volume  = {378},
  number  = {6617},
  pages   = {270--276},
  year    = {2022},
  doi = {10.1126/science.abq8271}
}

@article{Biamonte2017,
  author  = {Biamonte, Jacob and Wittek, Peter and Pancotti, Nicola and Rebentrost, Patrick and Wiebe, Nathan and Lloyd, Seth},
  title   = {Quantum machine learning},
  journal = {Nature},
  volume  = {549},
  number  = {7671},
  pages   = {195--202},
  year    = {2017},
  doi = {10.1038/nature23474}
}

@inproceedings{Strubell2019,
  author    = {Strubell, Emma and Ganesh, Ananya and McCallum, Andrew},
  title     = {Energy and Policy Considerations for Deep Learning in {NLP}},
  booktitle = {Proceedings of the 57th Annual Meeting of the Association for Computational Linguistics},
  publisher = {Association for Computational Linguistics},
  pages     = {3645--3650},
  year      = {2019},
  doi = {10.18653/v1/p19-1355}
}

@article{Barnett2009,
  author  = {Barnett, Stephen M. and Croke, Sarah},
  title   = {Quantum state discrimination},
  journal = {Advances in Optics and Photonics},
  volume  = {1},
  number  = {2},
  pages   = {238},
  year    = {2009},
  doi = {10.1364/aop.1.000238}
}

@article{Eldar2003,
  author  = {Eldar, Yonina C.},
  title   = {Mixed-quantum-state detection with inconclusive results},
  journal = {Physical Review A},
  volume  = {67},
  number  = {4},
  pages   = {042309},
  year    = {2003},
  doi = {10.1103/physreva.67.042309}
}

@article{Zernike1942,
  author  = {Zernike, F.},
  title   = {Phase contrast, a new method for the microscopic observation of transparent objects part {II}},
  journal = {Physica},
  volume  = {9},
  number  = {10},
  pages   = {974--986},
  year    = {1942},
  doi = {10.1016/s0031-8914(42)80079-8}
}

@article{Lecun1998,
  author  = {{LeCun}, Y. and Bottou, L. and Bengio, Y. and Haffner, P.},
  title   = {Gradient-based learning applied to document recognition},
  journal = {Proceedings of the IEEE},
  volume  = {86},
  number  = {11},
  pages   = {2278--2324},
  year    = {1998},
  doi = {10.1109/5.726791}
}

@article{Zhang2019,
  author  = {Zhang, He and Kumar, Santosh and Huang, Yu-Ping},
  title   = {Mode Selective Up-conversion Detection with Turbulence},
  journal = {Scientific Reports},
  volume  = {9},
  number  = {1},
  pages   = {17481},
  year    = {2019},
  doi = {10.1038/s41598-019-53914-8}
}

@article{Li2024b,
  author  = {Li, Lili and Kumar, Santosh and Sua, Yong Meng and Huang, Yu-Ping},
  title   = {Noise-resilient single-pixel compressive sensing with single photon counting},
  journal = {Communications Physics},
  volume  = {7},
  number  = {1},
  pages   = {110},
  year    = {2024},
  doi = {10.1038/s42005-024-01603-y}
}

@article{Cheng2022,
  author  = {Cheng, Yu and Zhao, Xin-Yu and Li, Li-Jing and Sun, Ming-Jie},
  title   = {First-photon imaging with independent depth reconstruction},
  journal = {APL Photonics},
  volume  = {7},
  number  = {3},
  pages   = {036103},
  year    = {2022},
  doi = {10.1063/5.0086159}
}

@article{Wu2026,
  author  = {Wu, Zi-Nan and Qin, Yu-Hao and Shi, Chao-Feng and Song, Jing-Ming and Deng, Wei-Jie and Sun, Ming-Jie},
  title   = {First-photon imaging denoising based on point cloud radius filtering},
  journal = {Optics and Lasers in Engineering},
  volume  = {196},
  pages   = {109420},
  year    = {2026},
  doi = {10.1016/j.optlaseng.2025.109420}
}

@article{Li2022,
  author  = {Li, Zhupeng and Liu, Xintong and Wang, Jianyu and Shi, Zuoqiang and Qiu, Lingyun and Fu, Xing},
  title   = {Fast non-line-of-sight imaging based on first photon event stamping},
  journal = {Optics Letters},
  volume  = {47},
  number  = {8},
  pages   = {1928},
  year    = {2022},
  doi = {10.1364/ol.446079}
}

@article{Ma2025,
  author  = {Ma, Shi-Yuan and Wang, Tianyu and Laydevant, Jérémie and Wright, Logan G. and McMahon, Peter L.},
  title   = {Quantum-limited stochastic optical neural networks operating at a few quanta per activation},
  journal = {Nature Communications},
  volume  = {16},
  number  = {1},
  pages   = {359},
  year    = {2025},
  doi = {10.1038/s41467-024-55220-y}
}

@article{Wang2023,
  author  = {Wang, Tianyu and Sohoni, Mandar M. and Wright, Logan G. and Stein, Martin M. and Ma, Shi-Yuan and Onodera, Tatsuhiro and Anderson, Maxwell G. and McMahon, Peter L.},
  title   = {Image sensing with multilayer nonlinear optical neural networks},
  journal = {Nature Photonics},
  volume  = {17},
  number  = {5},
  pages   = {408--415},
  year    = {2023},
  doi = {10.1038/s41566-023-01170-8}
}

@article{Kirmani2014,
  author  = {Kirmani, Ahmed and Venkatraman, Dheera and Shin, Dongeek and Colaço, Andrea and Wong, Franco N. C. and Shapiro, Jeffrey H. and Goyal, Vivek K},
  title   = {First-Photon Imaging},
  journal = {Science},
  volume  = {343},
  number  = {6166},
  pages   = {58--61},
  year    = {2014},
  doi = {10.1126/science.1246775}
}

@article{Naruse2016,
  author  = {Naruse, Makoto and Berthel, Martin and Drezet, Aurélien and Huant, Serge and Hori, Hirokazu and Kim, Song-Ju},
  title   = {Single Photon in Hierarchical Architecture for Physical Decision Making: Photon Intelligence},
  journal = {ACS Photonics},
  volume  = {3},
  number  = {12},
  pages   = {2505--2514},
  year    = {2016},
  doi = {10.1021/acsphotonics.6b00742}
}

@article{Gayer2008,
  author  = {Gayer, O. and Sacks, Z. and Galun, E. and Arie, A.},
  title   = {Temperature and wavelength dependent refractive index equations for {MgO}-doped congruent and stoichiometric {LiNbO$_3$}},
  journal = {Applied Physics B},
  volume  = {91},
  pages   = {343--348},
  year    = {2008},
  doi = {10.1007/s00340-008-2998-2}
}

@book{Helstrom1976,
  author    = {Helstrom, Carl W.},
  title     = {Quantum Detection and Estimation Theory},
  publisher = {Academic Press},
  address   = {New York},
  year      = {1976}
}

@article{Qiu2019,
  author  = {Qiu, Xiaodong and Zhang, Dongkai and Zhang, Wuhong and Chen, Lixiang},
  title   = {Structured-Pump-Enabled Quantum Pattern Recognition},
  journal = {Physical Review Letters},
  volume  = {122},
  number  = {12},
  pages   = {123901},
  year    = {2019},
  doi = {10.1103/physrevlett.122.123901}
}

@article{Ortolano2023,
  author  = {Ortolano, Giuseppe and Napoli, Carmine and Harney, Cillian and Pirandola, Stefano and Leonetti, Giuseppe and Boucher, Pauline and Losero, Elena and Genovese, Marco and Ruo-Berchera, Ivano},
  title   = {Quantum-Enhanced Pattern Recognition},
  journal = {Physical Review Applied},
  volume  = {20},
  number  = {2},
  pages   = {024072},
  year    = {2023},
  doi = {10.1103/physrevapplied.20.024072}
}

@article{Labroille2014,
  author  = {Labroille, Guillaume and Denolle, Bertrand and Jian, Pu and Genevaux, Philippe and Treps, Nicolas and Morizur, Jean-François},
  title   = {Efficient and mode selective spatial mode multiplexer based on multi-plane light conversion},
  journal = {Optics Express},
  volume  = {22},
  number  = {13},
  pages   = {15599},
  year    = {2014},
  doi = {10.1364/oe.22.015599}
}

@article{Fontaine2019,
  author  = {Fontaine, Nicolas K. and Ryf, Roland and Chen, Haoshuo and Neilson, David T. and Kim, Kwangwoong and Carpenter, Joel},
  title   = {{Laguerre-Gaussian} mode sorter},
  journal = {Nature Communications},
  volume  = {10},
  number  = {1},
  pages   = {1865},
  year    = {2019},
  doi = {10.1038/s41467-019-09840-4}
}

@article{Arrizon2007,
  author  = {Arrizón, Victor and Ruiz, Ulises and Carrada, Rosibel and González, Luis A.},
  title   = {Pixelated phase computer holograms for the accurate encoding of scalar complex fields},
  journal = {Journal of the Optical Society of America A},
  volume  = {24},
  number  = {11},
  pages   = {3500},
  year    = {2007},
  doi = {10.1364/josaa.24.003500}
}

@article{Ma2020,
  author  = {Ma, Zhaohui and Chen, Jia-Yang and Li, Zhan and Tang, Chao and Sua, Yong Meng and Fan, Heng and Huang, Yu-Ping},
  title   = {Ultrabright Quantum Photon Sources on Chip},
  journal = {Physical Review Letters},
  volume  = {125},
  number  = {26},
  pages   = {263602},
  year    = {2020},
  doi = {10.1103/physrevlett.125.263602}
}

@article{Erhard2018,
  author  = {Erhard, Manuel and Fickler, Robert and Krenn, Mario and Zeilinger, Anton},
  title   = {Twisted photons: new quantum perspectives in high dimensions},
  journal = {Light: Science \& Applications},
  volume  = {7},
  number  = {3},
  pages   = {17146},
  year    = {2018},
  doi = {10.1038/lsa.2017.146}
}

@misc{Zhang2023,
  author  = {Zhang, Yuanhang and Fontaine, Nicolas K.},
  title   = {Multi-Plane Light Conversion: A Practical Tutorial},
  year    = {2023},
  eprint = {2304.11323},
  archivePrefix = {arXiv}
}

@inproceedings{Kingma2015,
  author    = {Kingma, Diederik P. and Ba, Jimmy},
  title     = {Adam: A Method for Stochastic Optimization},
  booktitle = {3rd International Conference on Learning Representations (ICLR)},
  year      = {2015},
  eprint = {1412.6980},
  archivePrefix = {arXiv}
}

@misc{Fan2026,
  author = {Fan, Yanshan and Hu, Jianyong and Wu, Shuxiao and Qiao, Zhixing and Feng, Guosheng and Yang, Changgang and Liu, Jianqiang and Chen, Ruiyun and Qin, Chengbing and Zhang, Guofeng and Xiao, Liantuan and Jia, Suotang},
  title  = {Quantum compressed sensing enables image classification with a single photon},
  year   = {2026},
  eprint = {2604.25480},
  archivePrefix = {arXiv}
}

@misc{Ma2026,
  author = {Ma, Shi-Yuan and Laydevant, J{\'e}r{\'e}mie and Sohoni, Mandar M. and Wright, Logan G. and Wang, Tianyu and McMahon, Peter L.},
  title  = {Machine vision with small numbers of detected photons per inference},
  year   = {2026},
  eprint = {2603.23974},
  archivePrefix = {arXiv}
}

@misc{Duplinskiy2026,
  author = {Duplinskiy, A. and Filipovich, M. J. and Lvovsky, A. I.},
  title  = {Experimental visual quantum state discrimination},
  year   = {2026},
  eprint = {2611.10282},
  archivePrefix = {arXiv}
}

@misc{Sohoni2026,
  author = {Sohoni, Mandar M. and Laydevant, J{\'e}r{\'e}mie and Ouellet, Mathieu and Ma, Shi-Yuan and Yanagimoto, Ryotatsu and Ash, Benjamin A. and Onodera, Tatsuhiro and Wang, Tianyu and Wright, Logan G. and McMahon, Peter L.},
  title  = {Ultra-low-light computer vision using trained photon correlations},
  year   = {2026},
  eprint = {2604.11993},
  archivePrefix = {arXiv}
}

@article{Anderson2024,
  author  = {Anderson, Maxwell G. and Ma, Shi-Yuan and Wang, Tianyu and Wright, Logan G. and McMahon, Peter L.},
  title   = {Optical Transformers},
  journal = {Transactions on Machine Learning Research},
  year    = {2024},
  eprint  = {2302.10360},
  archivePrefix = {arXiv}
}

@article{Darji2024,
  author  = {Darji, Ishan and Kumar, Santosh and Huang, Yu-Ping},
  title   = {Robust super-resolution classifier by nonlinear optics},
  journal = {Optics Letters},
  volume  = {49},
  number  = {19},
  pages   = {5419--5422},
  year    = {2024},
  doi = {10.1364/OL.537295}
}

@article{Zhang2020,
  author  = {Zhang, He and Kumar, Santosh and Huang, Yu-Ping},
  title   = {Super-resolution optical classifier with high photon efficiency},
  journal = {Optics Letters},
  volume  = {45},
  number  = {18},
  pages   = {4968--4971},
  year    = {2020},
  doi = {10.1364/OL.401614}
}

@misc{NLONNrepo,
  author       = {Darji, Ishan},
  title        = {Data and code for ``{All-Optical Neural Network by First-Photon Detection}''},
  year         = {2026},
  howpublished = {\url{https://github.com/idarji-stevens-edu/NLONN-first-photon}}
}

\end{document}